\documentclass{article} 
\usepackage{iclr2027_conference,times}

\usepackage{amsmath,amsfonts,bm}

\def\eqref#1{equation~\ref{#1}}

\def\1{\bm{1}}

\DeclareMathAlphabet{\mathsfit}{\encodingdefault}{\sfdefault}{m}{sl}
\SetMathAlphabet{\mathsfit}{bold}{\encodingdefault}{\sfdefault}{bx}{n}

\usepackage{hyperref}
\usepackage{url}
\usepackage{graphicx}
\usepackage{array}
\usepackage{booktabs}
\usepackage{amssymb}
\usepackage{multirow}
\usepackage{makecell}
\usepackage{tabularx}
\usepackage{wrapfig}
\usepackage[most]{tcolorbox}
\usepackage{tikz}
\usetikzlibrary{arrows.meta,positioning}
\usepackage{xcolor}
\usepackage{booktabs,multirow,makecell,graphicx,wrapfig}
\usepackage[table]{xcolor}
\definecolor{basebg}{RGB}{241,243,246}

\title{Learning When and How to Intervene: A Hindsight-Distilled Sentinel for Coding Agents}

\author{
\textbf{Jiangrui Zhao\textsuperscript{1},
Chenglong Li\textsuperscript{2},
Meng Zhang\textsuperscript{3},
Xiaoting Du\textsuperscript{1}\thanks{Corresponding author.}} \\
{\normalfont
\textsuperscript{1}Beijing University of Posts and Telecommunications} \\
{\normalfont
\textsuperscript{2}Beijing University of Technology \quad
\textsuperscript{3}Meta}
}

\iclrfinalcopy 
\begin{document}

\maketitle
\fancyhead{}                       
\renewcommand{\headrulewidth}{0pt} 

\begin{abstract} 
Coding agents solve repository-level tasks through sequences of actions, where a single erroneous action can misdirect subsequent decisions and increase recovery costs.
Existing approaches use execution feedback for recovery or specialized checks to block errors, but deciding before execution whether intervention will benefit eventual task completion remains challenging.
To address this challenge, we propose \textsc{HiSentinel}, a hindsight-distillation framework that trains lightweight 0.6B and 1.7B sentinels to select pre-execution interventions aimed at improving task completion rather than correcting every imperfect action.
A privileged teacher uses recorded execution outcomes as evidence for intervention judgments, which are distilled into a causal student that receives only the pre-action context and proposed action.
Beyond identifying whether and when to intervene, the sentinel must also provide actionable feedback that helps the coding agent recover or obtain necessary human input.
To support these capabilities, we introduce \textsc{SWE-Intervene}, an action-level dataset constructed from software-engineering trajectories that annotates whether an action should be allowed, autonomously redirected, or paused for human assistance, together with corresponding intervention feedback.
Across SWE-bench Verified Mini and Ask or Assume, \textsc{HiSentinel} consistently improves task completion across Sentinel scales and coding-agent families, with gains of up to 14\% and 10\%, respectively, while maintaining competitive token consumption. These results demonstrate that lightweight pre-execution intervention can effectively prevent error propagation and improve the reliability of autonomous coding agents.
\end{abstract}

\section{Introduction}
Large language models now power coding agents that solve repository-level software-engineering tasks by navigating codebases, editing files, executing commands, and running tests~\citep{jimenez2024swe,yang2024swe,wang2025openhands}. Because these agents operate through multi-step interaction loops, a poor action can alter the environment, mislead subsequent decisions, and propagate into failed repairs, redundant exploration, or ultimately task failure~\citep{chen2025beyond,majgaonkar2025understanding,gandhi2025agents}. Critically, an early mistake may remain latent for several steps and become apparent only after recovery has become difficult or impossible~\citep{zhao2026failure}.

Evidence from Terminal-Bench highlights this gap: among 1,184 failed CLI-agent trajectories, half committed their decisive error by step 7 and became unrecoverable around step 12, while the first observable failure appeared only around step 16; prefix monitoring detected just 3.7--8.7\% of failures before lock-in~\citep{merrill2026terminal, zhao2026failure}. Existing process critics review recent steps and provide periodic corrective feedback~\citep{gandhi2025agents}, while trajectory diagnosis can locate decisive errors after a run~\citep{wang2026trajaudit,zhang2026longrca}. Yet feedback may arrive after a harmful action has executed, and retrospective diagnosis cannot prevent it. This leaves open how to judge an action when it is first proposed.

\begin{figure}[t]
    \centering
    \includegraphics[width=\columnwidth]{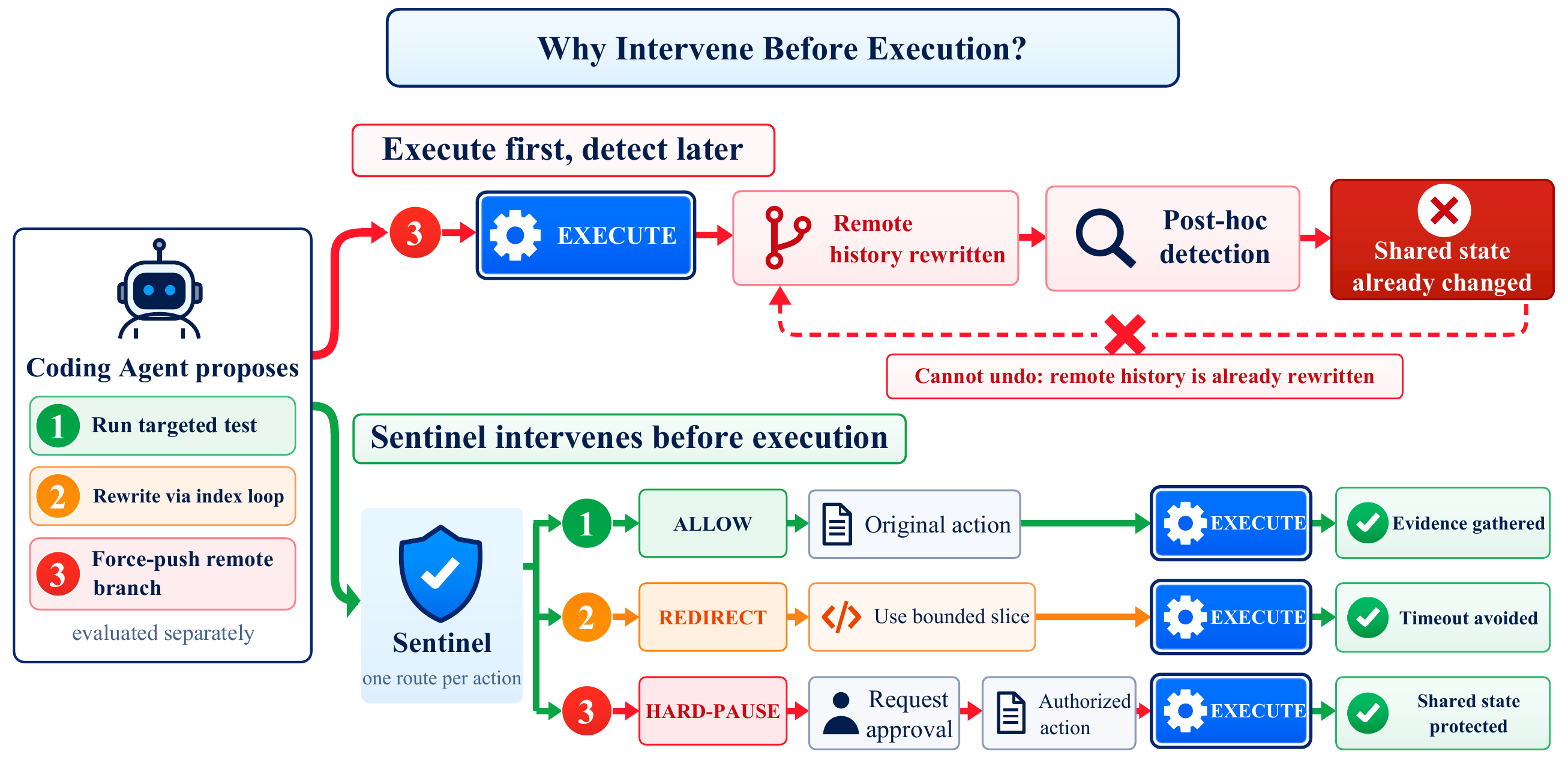}
    \caption{Pre-execution intervention for coding agents. The Sentinel evaluates each proposed action before execution and selects \textsc{Allow}, \textsc{Redirect}, or \textsc{Hard-Pause}.}
    \label{fig:overview}
    \vspace{-2em}
\end{figure}

As illustrated in Figure~\ref{fig:overview}, this motivates moving intervention earlier: after the agent proposes an action but before the environment executes it. At this boundary, a Sentinel can prevent a harmful action from changing the environment, rather than attempting to recover from its consequences afterward. We formulate this as a three-way decision among \textsc{Allow}, \textsc{Redirect}, and \textsc{Hard-Pause}: allowing appropriate actions, redirecting errors the agent can correct autonomously, and pausing when progress requires information, authorization, or access available only from a human.

However, moving intervention before execution creates a fundamental information gap. The execution result provides the most direct evidence of whether a proposed action is appropriate, yet this evidence is inherently unavailable at the time of intervention. The Sentinel must therefore anticipate whether an action threatens eventual task completion using only the pre-execution context and the proposed action.

To bridge this gap, we propose \textsc{HiSentinel}, a hindsight-distilled framework that transfers knowledge available after execution into a model that acts before execution. During training, a privileged teacher observes both the proposed action and its recorded execution result, allowing it to judge the action using evidence of what actually occurred. Its judgments are then distilled into a lightweight causal Sentinel that receives only the pre-execution context and proposed action. At deployment, the Sentinel can therefore intervene without access to future observations while benefiting from supervision informed by execution hindsight.

Beyond deciding whether to intervene, the Sentinel must communicate how the agent should proceed. We therefore optimize feedback generation separately, training the Sentinel to provide concise, evidence-grounded guidance that suggests a minimal correction or requests the specific human input needed to continue.

To train both intervention decisions and feedback, we introduce \textsc{SWE-Intervene}, an action-level dataset constructed by selecting decision nodes from natural software-engineering trajectories. Each node preserves the task, complete pre-action context, and exact proposed action, and is annotated as a three-way decision.

We evaluate \textsc{HiSentinel} at two levels: intervention decisions and feedback quality on frozen in-domain tests and external trajectory-diagnosis benchmarks, and end-to-end performance with coding agents on \textsc{SWE-bench Verified mini} and \textsc{Ask-or-Assume}. We measure task completion, unnecessary interventions, and inference overhead~\citep{hobbhahn2025swebenchmini,edwards2026ask}.

In summary, our contributions are as follows:
\begin{itemize}
    \item We formulate real-time intervention at the boundary between action proposal and execution as a three-way decision among \textsc{Allow}, \textsc{Redirect}, and \textsc{Hard-Pause}, targeting eventual task completion rather than merely detecting imperfect actions.

    \item We introduce \textsc{SWE-Intervene}, an action-level dataset for learning when and how to intervene in natural software-engineering trajectories.

    \item We propose \textsc{HiSentinel}, which distills execution hindsight from a privileged teacher into a lightweight pre-execution Sentinel and trains it to generate concise corrective feedback that guides subsequent agent behavior.
\end{itemize}

\section{Related Work}

\paragraph{Trajectory-based Agent Analysis.}
Agent evaluation has evolved from final task success and progress-based metrics in \textsc{AgentBoard}~\citep{ma2024agentboard} to trace-level failure attribution and localization in \textsc{Who\&When} and \textsc{TRAIL}~\citep{zhang2025agent,deshpande2025trail}. Recent methods further analyze realistic trajectories through learned or structured diagnosis, including \textsc{AgenTracer}, \textsc{TrajAudit}, \textsc{FALAT}, \textsc{TrajDebug}, and \textsc{LongRCA}~\citep{zhang2026agentracer,ou2025agentdiagnose,liu2026process,barke2026agentrx,wang2026trajaudit,rafi2026falat,shu2026resolve,zhao2026failure,qi2026trajdebug,zhang2026longrca}. Unlike these primarily diagnostic approaches, \textsc{HiSentinel} asks whether intervention at the current node has sufficient evidence to improve eventual task completion over autonomous recovery.

\paragraph{Learning with Privileged Information.}
Learning with privileged information transfers training-only evidence from an informed teacher to a restricted student~\citep{vapnik2015learning,chen2020learning,cai2024provable}. Recent work extends this paradigm to LLMs: \textsc{LEAP} uses privileged environment states to provide corrective feedback for LLM agents~\citep{choudhury2025better}, while \textsc{OPSD} conditions a self-teacher on verified solutions and $\pi$-\textsc{Distill} jointly optimizes privileged and unprivileged policies~\citep{zhao2026self,penaloza2026privileged}. Related methods exploit future-conditioned, short-context, or trajectory-derived privilege~\citep{fang2026privileged,zhang2026opsdl,dat2026dopsd,chen2026dualopsd}, although privileged context can be harmful when unrealizable from the student view~\citep{kaur2026rethinking,shrestha2026rethinking}. In contrast, \textsc{HiSentinel} uses realized software-engineering continuations as privileged evidence for intervention utility, rather than supervising task actions or reasoning trajectories.

\section{\textsc{SWE-Intervene}: Dataset Construction}
\label{sec:dataset}

\textsc{SWE-Intervene} draws action-level instances from \textsc{Open-SWE-Traces}, \textsc{SWE-Hero}, and \textsc{SWE-chat}~\citep{ahmad2026open,ludwig2026swe,baumann2026swe}. Using human-defined guidelines, GPT-5.5~\citep{openai2026gpt55} labels decision points as \textsc{Allow}, \textsc{Redirect}, or \textsc{Hard-Pause}; \textsc{SWE-chat} \texttt{AskUserQuestion} (AUQ) interactions supply natural \textsc{Hard-Pause} cases. A two-stage human audit rejects 25.3\% of intervention candidates. The resulting data comprise 5,680 training and 1,243 held-out test instances (82.0\%/18.0\%); the test set includes 1,122 natural and 121 AUQ instances. We de-duplicate training data against every downstream benchmark in Section~\ref{sec:setup} using instance identities and normalized task, repository, and trajectory signatures. Appendix~\ref{app:data-audit} details the sources, audit, and de-duplication.

\section{\textsc{HiSentinel}}
\label{sec:method}

\begin{figure}[t]
    \centering
    \includegraphics[width=\textwidth]{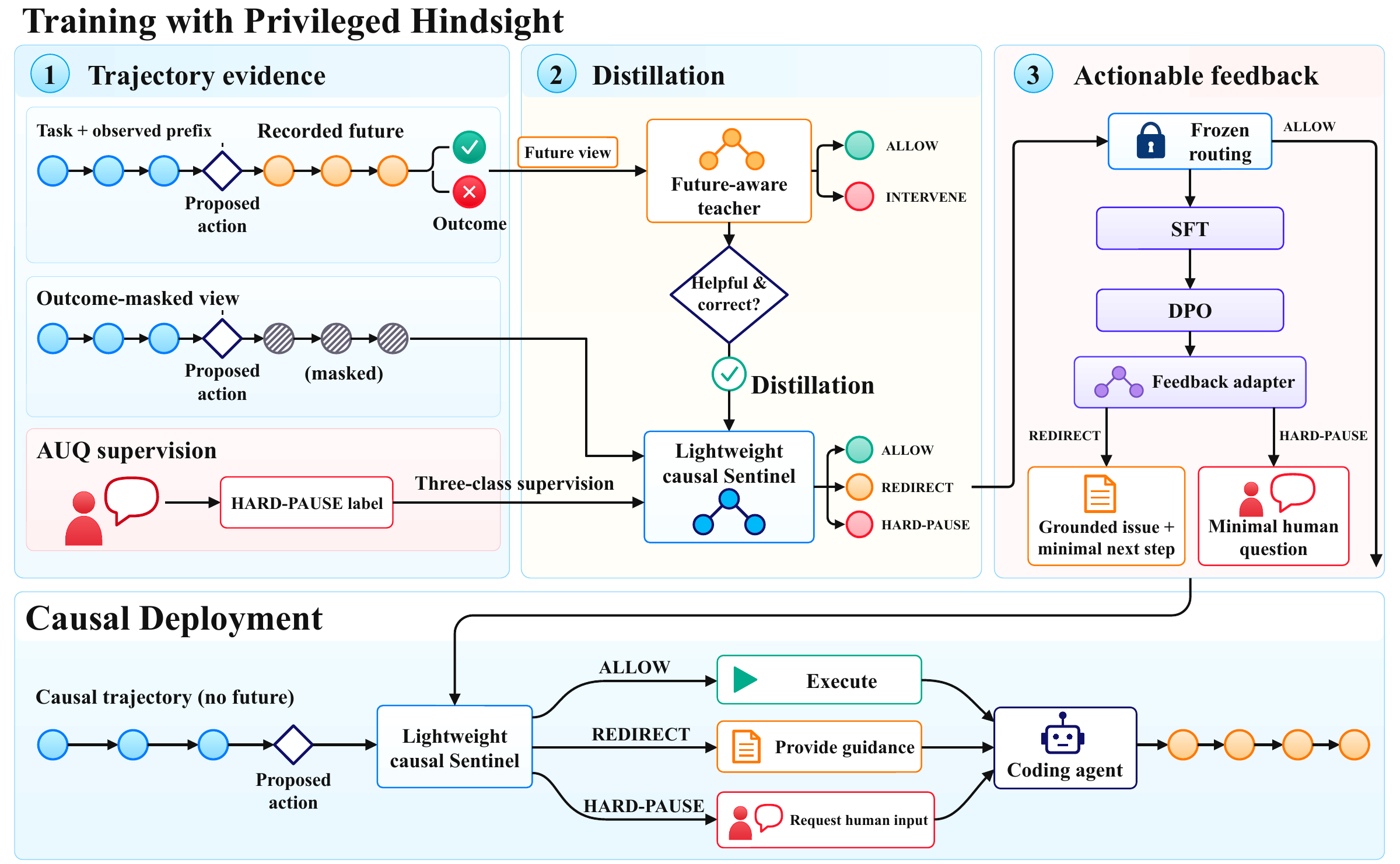}
    \caption{Overview of \textsc{HiSentinel}. Privileged hindsight is used only during training, while the deployed causal sentinel selects an intervention before the proposed action is executed.}
    \label{fig:hisentinel-overview}
    \vspace{-1.5em}
\end{figure}

Figure~\ref{fig:hisentinel-overview} presents the overall framework of \textsc{HiSentinel}. During training, a future-aware teacher uses recorded outcomes to provide privileged intervention supervision, which is distilled into a lightweight causal sentinel, while AUQ-derived samples supply supervision for \textsc{Hard-Pause}. A separate feedback adapter is optimized to generate actionable guidance or human-assistance requests. At deployment, the sentinel observes only the trajectory prefix and proposed action, and intervenes before execution when necessary.

\subsection{Problem Formulation}
\label{sec:problem-formulation}

We consider a coding agent solving a software-engineering task \(u\) through a sequence of actions and environment observations. At step \(t\), the agent has observed a trajectory prefix \(h_t\) and proposes an action \(a_t\). Before the action is executed, \textsc{HiSentinel} evaluates the pre-action context
\begin{equation}
    x_t = (u,h_t,a_t),
\end{equation}
and returns an intervention route \(\hat y_t\) together with feedback \(r_t\) when intervention is required. The route and feedback determine how the coding agent proceeds. Our objective is to intervene when there is sufficient reason to expect an improvement in eventual task completion relative to autonomous continuation, rather than to correct every locally imperfect action.

\begin{table}[h]
\caption{Intervention actions used by \textsc{HiSentinel}.}
\label{tab:intervention-actions}
\centering
\small
\setlength{\tabcolsep}{6pt}
\renewcommand{\arraystretch}{1.12}
\setlength{\aboverulesep}{4pt}
\setlength{\belowrulesep}{4pt}
\begin{tabular}{@{}
    >{\centering\arraybackslash}m{0.18\linewidth}
    >{\raggedright\arraybackslash}m{\dimexpr0.82\linewidth-2\tabcolsep\relax}
@{}}
\toprule
\textbf{Action} & \multicolumn{1}{c}{\textbf{Operational definition}} \\
\midrule
\textsc{Allow} & Execute the proposed action without intervention because there is insufficient evidence that intervening would improve task completion. \\
\midrule
\textsc{Redirect} & Withhold the proposed action and provide corrective feedback, prompting the agent to revise its plan and propose a new action autonomously. \\
\midrule
\textsc{Hard-Pause} & Suspend execution and request missing information, authorization, or preferences from a human before the agent continues. \\
\bottomrule
\end{tabular}
\vspace{-1.5em}
\end{table}

The three routes induce different runtime control flows. Under \textsc{Allow}, \(r_t=\varnothing\) and \(a_t\) is executed unchanged. Under \textsc{Redirect}, \(r_t\) provides corrective guidance; the original action is withheld, and the agent proposes a revised action conditioned on the feedback. Under \textsc{Hard-Pause}, \(r_t\) specifies a question for the human, and execution remains suspended until the answer is returned to the agent.

For training, let \(e_t\) denote the retrospective evidence available at a decision point, such as its recorded continuation and task outcome or an observed agent--human interaction. Following the annotation criteria in Section~\ref{sec:dataset}, this evidence informs the intervention label
\begin{equation}
    y_t=\mathcal{A}(x_t,e_t)\in\mathcal{Y},
    \qquad
    \mathcal{Y}=\{\textsc{Allow},\textsc{Redirect},\textsc{Hard-Pause}\}.
\end{equation}
The annotation assesses whether intervention has a concrete reason to help task completion, including whether the agent is already able to recover autonomously. Each logged decision point reveals only its realized continuation, so \(y_t\) is a hindsight-informed judgment rather than a measured difference between outcomes under alternative routes. \textsc{HiSentinel} learns \(p_{\theta}(y_t\mid x_t)\); at deployment, neither \(e_t\) nor any subsequent outcome is available.

\subsection{Hindsight-Guided Intervention Distillation}
\label{sec:hgid}

Our goal is to transfer the evidence contained in completed trajectories into a lightweight sentinel that must operate before action execution. We therefore introduce \textbf{H}indsight-\textbf{G}uided \textbf{I}ntervention \textbf{D}istillation (\textsc{HGID}). A privileged teacher observes the future and task outcome, while the causal student receives only the task, trajectory prefix, and proposed action. Distillation is applied only when the future improves the teacher's prediction over the corresponding outcome-masked view.

Because natural \textsc{Hard-Pause} events are scarce, the teacher predicts only \textsc{Allow} or \textsc{Intervene}. AUQ-derived \textsc{Hard-Pause} examples are excluded from teacher training because their suffixes occur after the missing information has already been supplied, but they remain available for three-class student supervision. To support subsequent feedback generation, we formulate routing as generative classification: the model predicts the route through its native language-modeling head rather than a separate classification head~\citep{raffel2020exploring,zhang2025generative,he2025gencls++}. The generated route can consequently serve as the prefix for producing corrective guidance or a request for human assistance.

\begin{equation}
    b_t = g(y_t) =
    \begin{cases}
        \textsc{Allow}, & y_t=\textsc{Allow},\\
        \textsc{Intervene}, &
        y_t\in\{\textsc{Redirect},\textsc{Hard-Pause}\}.
    \end{cases}
\end{equation}

Given the causal input \(x_t\) and its recorded future \(f_t\), the teacher is optimized by

\begin{equation}
    \mathcal{L}_{\mathrm{teacher}}
    =
    -\mathbb{E}_{(x_t,f_t,b_t)\sim\mathcal{D}_{F}}
    \log p_{\phi}(b_t\mid x_t,f_t),
\end{equation}

where \(\mathcal{D}_{F}\) contains trajectories with valid future evidence. The student retains the full three-class output space and is trained on the complete dataset, including AUQ-derived examples:

\begin{equation}
    \mathcal{L}_{\mathrm{route}}
    =
    -\mathbb{E}_{(x_t,y_t)\sim\mathcal{D}}
    \log p_{\theta}(y_t\mid x_t).
\end{equation}

Both models express their decisions through the native generative output space, without introducing a separate classification head. To align the binary teacher with the three-class student, we aggregate the student probabilities as

\begin{equation}
\begin{aligned}
    \bar{p}_{\theta}(\textsc{Allow}\mid x_t)
    &= p_{\theta}(\textsc{Allow}\mid x_t),\\
    \bar{p}_{\theta}(\textsc{Intervene}\mid x_t)
    &= p_{\theta}(\textsc{Redirect}\mid x_t)
     + p_{\theta}(\textsc{Hard-Pause}\mid x_t).
\end{aligned}
\end{equation}

This decomposition allows the teacher to supervise whether an intervention is warranted, while the three-class labels determine whether that intervention should be autonomous redirection or human assistance.

We use helpfulness gate to further prevent uninformative teacher predictions from dominating student learning. For each valid trajectory, we compare the teacher's future-aware prediction with a control view in which the privileged outcome is removed. Distillation is enabled only when the future-aware teacher predicts the correct binary label and assigns it a higher probability than the control view:

\begin{equation}
    m_t =
    \mathbb{I}\left[
        \arg\max_b p_{\phi}(b\mid x_t,f_t)=b_t
        \;\land\;
        p_{\phi}(b_t\mid x_t,f_t)
        >
        p_{\phi}(b_t\mid x_t,\varnothing)
    \right].
\end{equation}

The final student objective combines direct three-class supervision with selective future distillation:

\begin{equation}
\begin{aligned}
    \mathcal{L}_{\mathrm{HGID}}
    =\;&
    \mathcal{L}_{\mathrm{route}}\\
    &+
    \lambda\,
    \mathbb{E}_{(x_t,f_t)\sim\mathcal{D}_{F}}
    \left[
        m_t\tau^2
        D_{\mathrm{KL}}\!\left(
            p_{\phi}^{\tau}(\cdot\mid x_t,f_t)
            \,\Vert\,
            \bar{p}_{\theta}^{\tau}(\cdot\mid x_t)
        \right)
    \right],
\end{aligned}
\end{equation}

where \(\tau\) is the distillation temperature and \(\lambda\) controls the strength of privileged supervision. After training, the teacher and all future information are removed, leaving a lightweight three-class sentinel that operates solely on the causal pre-action context.

\subsection{From Intervention Decisions to Actionable Feedback}
\label{sec:actionable-feedback}

A route label alone does not specify how the agent should recover. We therefore activate a feedback module after the routing decision. For \textsc{Redirect}, it generates a grounded problem description and a minimal actionable suggestion; for \textsc{Hard-Pause}, it asks for the missing information, authorization, or preference. \textsc{Allow} requires no feedback.

We first perform supervised fine-tuning (SFT) on the annotated feedback in \textsc{SWE-Intervene}. Given the causal context $x_t$, fixed route $y_t$, and annotated response $r_t^{+}$, the feedback module is optimized with teacher forcing:
\begin{equation}
\mathcal{L}_{\mathrm{SFT}}
=
-\mathbb{E}
\left[
\frac{1}{|r_t^{+}|}
\sum_{k=1}^{|r_t^{+}|}
\log \pi_{\phi}
\left(
r_{t,k}^{+}
\mid x_t,y_t,r_{t,<k}^{+}
\right)
\right].
\end{equation}
This stage teaches the basic structure and content of intervention feedback.

Starting from the SFT policy, we further refine feedback quality using DPO~\citep{rafailov2023direct} over preferred and rejected responses $(r_t^{+},r_t^{-})$. Let
\begin{equation}
\begin{aligned}
\Delta_{\psi}
={}&
\log\frac{\pi_{\psi}(r_t^{+}\mid x_t,y_t)}
{\pi_{\mathrm{ref}}(r_t^{+}\mid x_t,y_t)}
-
\log\frac{\pi_{\psi}(r_t^{-}\mid x_t,y_t)}
{\pi_{\mathrm{ref}}(r_t^{-}\mid x_t,y_t)},
\end{aligned}
\end{equation}
where $\pi_{\mathrm{ref}}$ is a frozen copy of the SFT policy. We retain an NLL term on the preferred response while optimizing the preference objective:
\begin{equation}
\mathcal{L}_{\mathrm{feedback}}
=
\mathcal{L}_{\mathrm{NLL}}(r_t^{+})
-
\eta\,
\mathbb{E}
\left[
\log \sigma\!\left(\beta\Delta_{\psi}\right)
\right].
\end{equation}

Both SFT and DPO update only the feedback module. We freeze the routing parameters learned through \textsc{HGID} and condition feedback generation on the fixed predicted route. Consequently, feedback training cannot change whether the sentinel selects \textsc{Allow}, \textsc{Redirect}, or \textsc{Hard-Pause}; it only improves the response produced after an intervention is selected.

\section{Experiments}

\subsection{Experimental Setup}
\label{sec:setup}

\paragraph{Models.}
We use Qwen3-Coder-30B-A3B-Instruct as the privileged teacher and instantiate the lightweight sentinels with Qwen3-0.6B and Qwen3-1.7B~\citep{qwen3technicalreport}. At deployment, we use \textsc{mini-SWE-agent}~\citep{yang2024swe} as the coding-agent harness and evaluate the sentinels with two open-weight coding agents: Qwen3-Coder-30B-A3B-Instruct and Devstral-Small-2-24B-Instruct-2512~\citep{rastogi2025devstral}. The latter is a comparably sized model specialized for agentic software-engineering tasks, allowing us to evaluate whether the learned intervention policy transfers across coding-agent families. We additionally evaluate Claude-Sonnet-4.6~\citep{anthropic2026claudesonnet46} as a closed-source reference to assess transfer to a frontier proprietary coding agent.

\paragraph{Benchmarks.}
We evaluate \textbf{HiSentinel} on static intervention recognition and end-to-end task completion. For static recognition, \textbf{SWE-Intervene} evaluates three-way action classification among \textsc{Allow}, \textsc{Redirect}, and \textsc{Hard-Pause}; \textbf{RootSE}~\citep{wang2026trajaudit} localizes root causes in failed coding trajectories; and the \textsc{Program} subset of \textbf{R-Judge}~\citep{yuan2024r} classifies program-related reasoning trajectories as safe or unsafe. For end-to-end evaluation, \textbf{SWE-bench Verified Mini}~\citep{hobbhahn2025swebenchmini} is a 50-task subset of \textbf{SWE-bench Verified}~\citep{jimenez2024swe} preserving its performance, test-pass-rate, and difficulty distributions, with results closely matching the full benchmark across 16 models. \textbf{Ask or Assume}~\citep{edwards2026ask} masks task-critical details to test clarification behavior; we stratify 100 of its 500 tasks by repository and difficulty to approximate the full-set distribution. We cap each task at 30 minutes and 100 agent steps.

\paragraph{Baselines.}
For static intervention recognition, \textbf{Step-by-Step} makes predictions from the causal prefix by examining actions sequentially. Post-hoc baselines observe the complete trajectory: \textbf{All-at-Once} diagnoses it in a single prompt; \textbf{AgentRx} identifies failures by synthesizing and checking constraints in multi-agent trajectories~\citep{barke2026agentrx}; \textbf{TrajAudit} uses an investigator agent with context reduction to locate decisive errors~\citep{wang2026trajaudit}; and \textbf{RCTA} traces candidate errors through segment summaries, dependencies, and agent handoffs~\citep{zhang2026longrca}.


\begin{wraptable}{r}{0.64\textwidth}
\centering
\caption{
\textbf{End-to-end results.}
SWE-V Mini denotes SWE-bench Verified Mini;
SDS denotes \textsc{Steer, Don't Solve}.
Res.: resolved (\%).
Cost: average input and output tokens per trajectory (millions).
}
\label{tab:e2e-results}
\scriptsize
\setlength{\tabcolsep}{2.5pt}
\renewcommand{\arraystretch}{1.04}
\resizebox{\linewidth}{!}{%
\begin{tabular}{@{}cclcccc@{}}
\toprule
\multirow{2}{*}{\bfseries Agent} &
\multirow{2}{*}{\bfseries Backbone} &
\multirow{2}{*}{\bfseries Method} &
\multicolumn{2}{c}{\bfseries SWE-V Mini} &
\multicolumn{2}{c}{\bfseries Ask or Assume} \\
\cmidrule(lr){4-5}
\cmidrule(lr){6-7}
& & &
Res.~$\uparrow$ & Cost~$\downarrow$ &
Res.~$\uparrow$ & Cost~$\downarrow$ \\
\midrule

\multirow{2}{*}{\makecell{Sonnet-\\4.6}}
& --
& \cellcolor{basebg}Base
& \cellcolor{basebg}60.00 & \cellcolor{basebg}0.26
& \cellcolor{basebg}58.00 & \cellcolor{basebg}0.47 \\
& Qwen3-1.7B
& \textsc{HiSentinel}
& \textbf{66.00} & 0.41
& \textbf{61.00} & 0.82 \\
\midrule

\multirow{9}{*}{\makecell{Qwen3-\\Coder}}
& \multirow{2}{*}{--}
& \cellcolor{basebg}Base
& \cellcolor{basebg}30.00 & \cellcolor{basebg}0.29
& \cellcolor{basebg}23.00 & \cellcolor{basebg}0.26 \\
& & Reflexion
& 34.00 & 0.31 & 20.00 & 0.36 \\
\cmidrule(lr){2-7}
& SDS-4B & SDS
& 28.00 & 0.57 & 24.00 & 0.44 \\
\cmidrule(lr){2-7}
& \multirow{3}{*}{Qwen3-0.6B}
& Step-by-Step
& 18.00 & 0.12 & 20.00 & 0.30 \\
& & SWE-PRM
& 32.00 & 0.38 & 22.00 & 0.32 \\
& & \textsc{HiSentinel}
& \textbf{36.00} & 0.38
& \textbf{28.00} & 0.34 \\
\cmidrule(lr){2-7}
& \multirow{3}{*}{Qwen3-1.7B}
& Step-by-Step
& 32.00 & 0.45 & 21.00 & 0.33 \\
& & SWE-PRM
& 20.00 & 0.35 & 25.00 & 0.36 \\
& & \textsc{HiSentinel}
& \textbf{44.00} & 0.43
& \textbf{33.00} & 0.39 \\
\midrule

\multirow{11}{*}{Devstral}
& \multirow{2}{*}{--}
& \cellcolor{basebg}Base
& \cellcolor{basebg}20.00 & \cellcolor{basebg}1.16
& \cellcolor{basebg}19.00 & \cellcolor{basebg}1.25 \\
& & Reflexion
& 24.00 & 1.31 & 21.00 & 1.40 \\
\cmidrule(lr){2-7}
& SDS-4B & SDS
& 24.00 & 1.46 & 20.00 & 1.53 \\
\cmidrule(lr){2-7}
& \multirow{3}{*}{Qwen3-0.6B}
& Step-by-Step
& 18.00 & 1.21 & 18.00 & 1.31 \\
& & SWE-PRM
& 22.00 & 1.28 & 21.00 & 1.35 \\
& & \textsc{HiSentinel}
& \textbf{26.00} & 1.30
& \textbf{24.00} & 1.39 \\
\cmidrule(lr){2-7}
& \multirow{3}{*}{Qwen3-1.7B}
& Step-by-Step
& 22.00 & 1.25 & 20.00 & 1.34 \\
& & SWE-PRM
& 24.00 & 1.30 & 22.00 & 1.39 \\
& & \textsc{HiSentinel}
& \textbf{30.00} & 1.34
& \textbf{27.00} & 1.44 \\
\bottomrule
\end{tabular}%
}
\vspace{-2em}
\end{wraptable}

For end-to-end evaluation, \textbf{Step-by-Step} is a prompt-only baseline that inspects each proposed action before execution. \textbf{SWE-PRM} provides taxonomy-guided corrective feedback every five executed steps~\citep{gandhi2025agents}, while \textbf{Steer, Don't Solve} uses an SFT- and DPO-trained critic to provide high-level guidance at the same interval~\citep{gandhi2026steer}. Because its DPO training data are not publicly available, we directly use the strongest publicly released checkpoint.

\subsection{Experiment Results}

\begin{table}[t]
\caption{
\textbf{Results on static intervention-recognition benchmarks.}
No data from RootSE or R-Judge are used for training.
M-F1 and I-F1 denote Macro-F1 and Intervention F1. BAcc denote Balanced accuracy. 
For RootSE, T@$k$ counts predictions within $k$ steps of the annotated earliest decisive error.
Cost denotes the average total number of input and output tokens per trajectory.
}
\label{tab:static-results}
\vspace{-0.5em}
\begin{center}
\setlength{\tabcolsep}{3.6pt}
\renewcommand{\arraystretch}{0.98}
\resizebox{\linewidth}{!}{
\begin{tabular}{llccccccccc}
\toprule

\multirow[c]{3}{*}{\bfseries Backbone}
&
\multirow[c]{3}{*}{\bfseries Method}
&
\multicolumn{3}{c}{\bfseries In-Domain}
&
\multicolumn{6}{c}{\bfseries OOD}
\\
\cmidrule(lr){3-5}
\cmidrule(lr){6-11}

&
&
\multicolumn{3}{c}{\bfseries SWE-Intervene}
&
\multicolumn{3}{c}{\bfseries RootSE}
&
\multicolumn{3}{c}{\bfseries R-Judge}
\\
\cmidrule(lr){3-5}
\cmidrule(lr){6-8}
\cmidrule(lr){9-11}

&
&
M-F1 $\uparrow$
&
I-F1 $\uparrow$
&
Cost $\downarrow$
&
T@0 $\uparrow$
&
T@1 $\uparrow$
&
Cost $\downarrow$
&
BAcc $\uparrow$
&
M-F1 $\uparrow$
&
Cost $\downarrow$
\\
\midrule

\multirow[c]{6}{*}{Qwen3-0.6B}
& Step-by-Step
& 27.23 & 21.21 & 8.08K
& 3.92 & 10.78 & 0.12M
& 60.68 & 60.61 & 0.58K
\\

& All-at-Once
& 28.88 & 29.79 & 8.08K
& 3.92 & 6.86 & 0.07M
& 60.02 & 60.77 & 0.66K
\\

& TrajAudit
& 30.23 & 19.13 & 10.16K
& 0.00 & 3.92 & 0.10M
& 50.66 & 34.92 & 1.80K
\\

& RCTA
& 28.61 & 19.75 & 11.28K
& 2.94 & 10.78 & 0.03M
& 50.00 & 32.09 & 1.30K
\\

& AgentRx
& 25.78 & 25.51 & 9.16K
& 0.98 & 7.84 & 2.28M
& 49.17 & 31.72 & 3.40K
\\

& \textbf{HiSentinel} (Ours)
& \textbf{79.99} & \textbf{75.22} & 5.91K
& \textbf{9.80} & \textbf{16.67} & 0.07M
& \textbf{65.09} & \textbf{61.68} & 1.71K
\\

\midrule

\multirow[c]{6}{*}{Qwen3-1.7B}
& Step-by-Step
& 29.10 & 36.46 & 8.10K
& 3.92 & 12.75 & 0.06M
& 52.76 & 35.99 & 0.37K
\\

& All-at-Once
& 27.31 & 31.38 & 8.10K
& 3.92 & 10.78 & 0.07M
& 57.48 & 56.04 & 0.66K
\\

& TrajAudit
& 29.94 & 38.42 & 10.26K
& 2.94 & 8.82 & 0.10M
& 57.48 & 55.77 & 1.30K
\\

& RCTA
& 27.94 & 35.89 & 11.42K
& 2.94 & 11.76 & 0.04M
& 49.61 & 36.89 & 1.29K
\\

& AgentRx
& 27.94 & 33.92 & 9.23K
& 2.94 & 12.74 & 3.09M
& \textbf{66.14} & \textbf{65.60} & 3.51K
\\

& \textbf{HiSentinel} (Ours)
& \textbf{81.77} & \textbf{78.97} & 5.91K
& \textbf{9.80} & \textbf{15.69} & 0.08M
& \textbf{66.14} & 65.09 & 1.24K
\\

\bottomrule
\end{tabular}
}
\end{center}

\vspace{-2.5em}
\end{table}

\paragraph{Static intervention recognition.}
\textsc{HiSentinel} substantially outperforms prompt-based and post-hoc baselines on the in-domain \textsc{SWE-Intervene} test set, while requiring fewer tokens per trajectory. The advantage also transfers to RootSE and R-Judge, neither of which is used for training: \textsc{HiSentinel} provides the strongest failure-localization results on RootSE and achieves the best or competitive classification performance on R-Judge. These results show that the learned intervention policy generalizes beyond the annotation distribution used for training.

\paragraph{End-to-end agent performance.}
More importantly, the improved intervention decisions translate into consistently higher task completion rates. Across Qwen3-Coder, Devstral, and Sonnet-4.6, both \textsc{HiSentinel} variants improve over the corresponding unassisted agents and competing intervention methods on SWE-bench Verified Mini and Ask or Assume. The largest gains are obtained with the 1.7B model: for Qwen3-Coder, it improves resolution from 30\% to 44\% on SWE-bench Verified Mini and from 23\% to 33\% on Ask or Assume. Similar improvements with Devstral and the stronger Sonnet-4.6 agent indicate that \textsc{HiSentinel} is not tied to a particular coding model. Although monitoring introduces additional token cost, the cost remains comparable to other feedback-based baselines while yielding substantially larger improvements in task completion.

\subsection{Ablation Study}

\vspace{-1em}

\paragraph{Ablation setup.}

\newsavebox{\AblationRoutingBox}

\begin{table}[h]
\centering
\caption{Ablations of \textsc{HiSentinel}.
(a) Routing on \textsc{SWE-Intervene}:
M-F1/I-F1 denote three-class/Intervention F1;
R-F1/HP-F1 denote class-wise F1.
(b) Feedback groundedness (Ground.) and actionability (Action.)
pass rates.
(c) Resolved rate and cost on SWE-bench Verified Mini.
F1, pass rates, and resolved rates are percentages.}
\label{tab:ablation}

\footnotesize
\setlength{\tabcolsep}{2.5pt}
\renewcommand{\arraystretch}{1.05}

\sbox{\AblationRoutingBox}{%
\begin{minipage}[t]{0.57\linewidth}
\vspace{0pt}
\begin{tabular*}{\linewidth}{
    @{\extracolsep{\fill}}lcccc@{}}
\toprule
\multicolumn{5}{@{}l}{
    \itshape (a) Intervention routing} \\
\addlinespace[2pt]
{\bfseries Variant}
& {\bfseries M-F1}
& {\bfseries I-F1}
& {\bfseries R-F1}
& {\bfseries HP-F1} \\
\midrule
w/ 3-Class Teacher
& 75.53 & 74.42 & 62.87 & 80.00 \\
w/ Classification Head
& 66.92 & 67.22 & 58.91 & 88.36 \\
w/o Distillation
& 76.42 & 73.31 & 60.30 & 86.74 \\
w/ Causal Teacher
& 77.16 & 75.08 & {\bfseries 68.22} & 84.63 \\
w/o Route SFT
& 68.91 & 61.84 & 54.73 & 84.26 \\
w/o Helpfulness Gate
& 78.72 & 77.41 & 61.06 & 91.18 \\
w/ Self-Distillation
& 76.30 & 74.46 & 61.21 & 88.67 \\
w/ Joint Feedback
& 78.67 & 77.96 & 57.30 & 92.49 \\
\addlinespace[2pt]
\textsc{HiSentinel} (Full)
& {\bfseries 81.77}
& {\bfseries 78.97}
& 66.30
& {\bfseries 92.90} \\
\bottomrule
\end{tabular*}
\end{minipage}%
}

\usebox{\AblationRoutingBox}%
\hfill
\begin{minipage}[t]
[\dimexpr\ht\AblationRoutingBox+\dp\AblationRoutingBox\relax]
[s]{0.41\linewidth}
\vspace{0pt}
\noindent
\begin{tabular*}{\linewidth}{
    @{\extracolsep{\fill}}lcc@{}}
\toprule
\multicolumn{3}{@{}l}{
    \itshape (b) Feedback quality} \\
{\bfseries Variant}
& {\bfseries Ground.}
& {\bfseries Action.} \\
\midrule
w/ Joint Feedback
& 66.9 & 63.2 \\
w/o DPO
& 69.4 & 65.8 \\
\textsc{HiSentinel} (Full)
& {\bfseries 76.8}
& {\bfseries 73.5} \\
\bottomrule
\end{tabular*}

\par\vfill

\noindent
\begin{tabular*}{\linewidth}{
    @{\extracolsep{\fill}}lcc@{}}
\toprule
\multicolumn{3}{@{}l}{
    \itshape (c) SWE-bench Verified Mini} \\
{\bfseries Variant}
& {\bfseries Resolved}$\uparrow$
& {\bfseries Cost}$\downarrow$ \\
\midrule
w/o Distillation
& 34.00 & 0.43 \\
w/ Causal Teacher
& 36.00 & 0.42 \\
\textsc{HiSentinel} (Full)
& \textbf{44.00} & 0.43 \\
\bottomrule
\end{tabular*}
\par\vspace{0pt} 
\end{minipage}
\vspace{-0.5em}
\end{table}

We ablate \textsc{HiSentinel} on teacher supervision, routing, and feedback learning. All variants use the 1.7B backbone with the same data and optimization settings. Due to the cost of interactive evaluation, we select two representative routing ablations for the end to end study.

\paragraph{Routing.}
On \textsc{SWE-Intervene}, the full model achieves 81.77 M-F1 and 78.97 I-F1. Ablating distillation, replacing the future-aware teacher with a causal or self-distilled alternative, or removing the helpfulness gate consistently degrades performance, supporting the contribution of each component in \textsc{HGID}. Alternative routing implementations, including a classification head and removing route SFT, lead to substantially larger drops.

\paragraph{Feedback learning.}
With routing held fixed, the full model reaches 76.8\% groundedness and 73.5\% actionability. Joint training lowers these rates to 66.9\% and 63.2\%, while removing DPO yields 69.4\% and 65.8\%, respectively.

\paragraph{End to end impact.}
On SWE-bench Verified Mini, removing distillation and using a causal teacher reduce resolution from 44\% to 34\% and 36\%, despite their smaller gaps in static F1 and nearly identical token costs. Both variants make fewer false alarms but also miss more necessary interventions. A coding agent can sometimes recognize a low-risk false alarm and continue productively; when an intervention is missed, the flawed action proceeds. This asymmetry makes intervention recall especially consequential during execution.

\subsection{Analysis}

\subsubsection{Does \textsc{HiSentinel} benefit from precise or frequent intervention?}
Effective oversight depends on intervening well, not often: mistimed feedback can derail an otherwise successful trajectory. We judge each intervention solely from the task, proposed action, and pre-intervention trajectory, since even rare stochastic continuations could make poor guidance appear successful; final rescues and regressions are reported separately. On \textsc{SWE-bench Verified Mini}, 109 of \textsc{HiSentinel}'s 242 interventions (45.0\%) are beneficial, 106 neutral, and 27 harmful; eight rescues and one regression raise the resolved count from 15/50 to 22/50. \textsc{Step-by-Step} has one beneficial intervention among 24 (4.2\%) and one rescue without regression; \textsc{SWE-PRM} has three among 284 (1.1\%), one rescue, and six regressions. Thus, \textsc{HiSentinel}'s gains reflect intervention quality rather than frequency.

\subsubsection{Why We Recompute the Causal Prefix}
Since each monitoring input largely extends the preceding causal prefix by one step, cross-step KV reuse appears to be a natural optimization. We evaluate per-task BF16 exact-prefix caching for the 1.7B \textsc{HiSentinel}, but do not adopt it in the final system. In the cached run, untruncated calls reuse 86.4\% of their input tokens; however, 906/1,869 calls require left truncation, for which reuse falls to 3.2\%, limiting the token-weighted overall reuse ratio to 29.8\%. Compared with full-prefix recomputation, the observed mean number of processed Sentinel tokens decreases only from 382K to 299K per task (22\%), while mean end-to-end time remains similar (757\,s vs.\ 763\,s). A matched audit of 896 token-identical routing calls further identifies 13 routing flips and three effective-intervention flips. On \textsc{SWE-bench Verified Mini}, these differences reduce resolution from 22/50 with full recomputation to 18/50 with cross-step caching. We therefore recompute the complete causal prefix at every monitoring step, while retaining ordinary within-generation KV caching.


\definecolor{casebadbg}{HTML}{FFF3F1}
\definecolor{casebadline}{HTML}{C94B40}
\definecolor{caseintbg}{HTML}{F5F1FF}
\definecolor{caseintline}{HTML}{7357B8}
\definecolor{casefixbg}{HTML}{EEF8F2}
\definecolor{casefixline}{HTML}{2D8A5B}
\definecolor{caseoutbg}{HTML}{F3F7FB}
\definecolor{caseoutline}{HTML}{527A9D}

\subsubsection{Case Study: Redirecting an Ineffective Patch}
\label{sec:case-study}

\begin{figure*}[h]
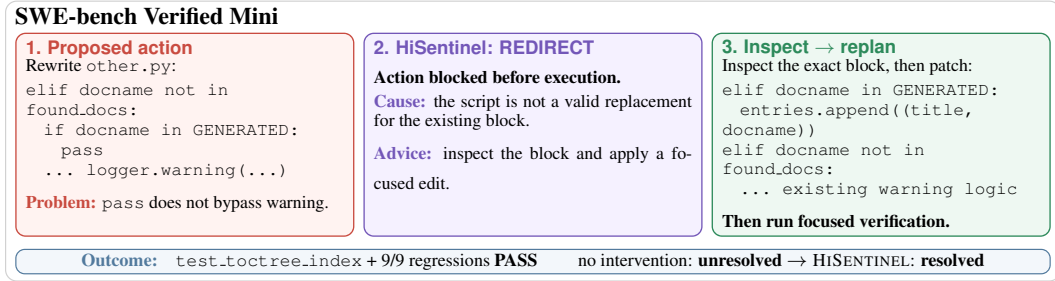

  \centering
  \begin{tcolorbox}[
      enhanced,
      colback=white,
      colframe=black!18,
      boxrule=0.45pt,
      arc=1.4mm,
      boxsep=0pt,
      left=1.2mm,right=1.2mm,top=0.8mm,bottom=0.8mm]

    {\footnotesize\bfseries SWE-bench Verified Mini}\par
    \vspace{0.7mm}

    \noindent\begin{tabularx}{\linewidth}{
        @{}X@{\hspace{1.2mm}}X@{\hspace{1.2mm}}X@{}}
      \begin{tcolorbox}[width=\linewidth,height=27mm,valign=top,
          colback=casebadbg,colframe=casebadline,
          boxrule=0.5pt,arc=1.2mm,
          boxsep=0pt,
          left=1.2mm,right=1.2mm,top=0.8mm,bottom=0.8mm]
        {\scriptsize\color{casebadline}\bfseries\sffamily
        1. Proposed action}\par\vspace{-0.3mm}
        {\fontsize{6.7}{7.5}\selectfont
        Rewrite \texttt{other.py}:\par
        \vspace{0.5mm}
        {\ttfamily
        elif docname not in found\_docs:\par
        \hspace*{1em}if docname in GENERATED:\par
        \hspace*{2em}pass\par
        \hspace*{1em}\ldots\ logger.warning(\ldots)}\par
        \vspace{0.5mm}
        {\color{casebadline}\bfseries Problem:}
        \texttt{pass} does not bypass warning.}
      \end{tcolorbox}
      &
      \begin{tcolorbox}[width=\linewidth,height=27mm,valign=top,
          colback=caseintbg,colframe=caseintline,
          boxrule=0.5pt,arc=1.2mm,
          boxsep=0pt,
          left=1.2mm,right=1.2mm,top=0.8mm,bottom=0.8mm]
        {\scriptsize\color{caseintline}\bfseries\sffamily
        2. HiSentinel: REDIRECT}\par\vspace{1.2mm}
        {\fontsize{6.7}{7.5}\selectfont
        \textbf{Action blocked before execution.}\par
        \vspace{0.5mm}
        {\color{caseintline}\bfseries Cause:}
        the script is not a valid replacement for the existing block.\par
        \vspace{0.5mm}
        {\color{caseintline}\bfseries Advice:}
        inspect the block and apply a focused edit.}
      \end{tcolorbox}
      &
      \begin{tcolorbox}[width=\linewidth,height=27mm,valign=top,
          colback=casefixbg,colframe=casefixline,
          boxrule=0.5pt,arc=1.2mm,
          boxsep=0pt,
          left=1.2mm,right=1.2mm,top=0.8mm,bottom=0.8mm]
        {\scriptsize\color{casefixline}\bfseries\sffamily
        3. Inspect $\rightarrow$ replan}\par\vspace{-0.3mm}
        {\fontsize{6.7}{7.5}\selectfont
        Inspect the exact block, then patch:\par
        \vspace{0.5mm}
        {\ttfamily
        elif docname in GENERATED:\par
        \hspace*{1em}entries.append((title, docname))\par
        elif docname not in found\_docs:\par
        \hspace*{1em}\ldots\ existing warning logic}\par
        \vspace{0.5mm}
        \textbf{Then run focused verification.}}
      \end{tcolorbox}
    \end{tabularx}

    \vspace{-1.5mm}
    \begin{tcolorbox}[
        colback=caseoutbg,colframe=caseoutline,
        boxrule=0.45pt,arc=1.1mm,
        boxsep=0pt,
        left=1mm,right=1mm,top=0.45mm,bottom=0.45mm]
      \centering\fontsize{6.8}{7.4}\selectfont
      {\color{caseoutline}\bfseries Outcome:}\quad
      \texttt{test\_toctree\_index} + 9/9 regressions \textbf{PASS}
      \qquad no intervention: \textbf{unresolved}
      $\rightarrow$ \textsc{HiSentinel}: \textbf{resolved}
    \end{tcolorbox}
  \end{tcolorbox}
  \vspace{-0.8em}
  \caption{A representative causal intervention.  The displayed
  \textsc{Redirect} is one local episode in the trajectory; the final
  resolved/unresolved labels are produced by the official SWE-bench evaluator.}
  \label{fig:redirect-case}
  \vspace{-0.8em}
\end{figure*}

Figure~\ref{fig:redirect-case} shows a representative \textsc{Redirect} episode from the 1.7B \textsc{HiSentinel} run on \texttt{sphinx-doc\_\_sphinx-10673}. The task requires Sphinx to suppress missing-document warnings for documents generated by extensions while preserving the warnings for genuinely missing documents. The coding agent attempted to implement this behavior with a nested branch containing an empty \texttt{pass}, which still allowed execution to reach the existing warning logic. \textsc{HiSentinel} blocked the edit and requested a focused, evidence-grounded correction. The agent then inspected the relevant block and moved the generated-document check before the missing-document branch. The resulting run passed \texttt{test\_toctree\_index} and all nine regression tests, whereas the paired no-intervention run remained unresolved. This episode illustrates how a \textsc{Redirect} steers the agent toward a concrete corrective action rather than merely providing a post-hoc warning.

\section{Conclusion}
We introduced \textsc{HiSentinel}, a pre-execution intervention framework that distills outcome-aware judgments from a privileged teacher into a lightweight causal Sentinel. Seeing only the observed trajectory and proposed action, it chooses among \textsc{Allow}, \textsc{Redirect}, and \textsc{Hard-Pause}, and provides actionable feedback when intervention is needed. We also introduced \textsc{SWE-Intervene} to study these decisions on natural coding trajectories. The 1.7B model achieves 81.77 macro-F1 and 78.97 intervention F1 on this benchmark and transfers to RootSE and R-Judge. In end-to-end evaluation with Qwen3-Coder, it improves resolution from 30\% to 44\% on SWE-bench Verified Mini and from 23\% to 33\% on Ask or Assume, with gains also observed for Devstral and Sonnet-4.6. These results show that hindsight supervision can improve coding agents through timely, useful intervention.

\subsection*{AI Use Statement}

Generative AI tools were used in this work to assist with language editing, the preparation and refinement of figures and tables, data annotation, and evaluation as LLM judges. All AI-assisted materials and annotations were reviewed and verified by the authors to ensure accuracy and faithful representation of the research. These tools were not used to independently produce experimental results, conduct scientific analyses, or formulate research conclusions. The authors retain full responsibility for all content presented in this paper.

\bibliography{iclr2027_conference}
\bibliographystyle{iclr2027_conference}

\appendix
\section{Appendix}

\subsection{\textsc{SWE-Intervene}: Construction and Quality Audit}
\label{app:data-audit}

\subsubsection{Data Sources and Instance Construction}

\textsc{SWE-Intervene} is constructed from three complementary sources of software-engineering agent trajectories. \textsc{Open-SWE-Traces}~\citep{ahmad2026open} provides repository-level trajectories generated by coding agents, while \textsc{SWE-Hero}~\citep{ludwig2026swe} contributes trajectories containing diverse development and debugging behaviors. We further use \textsc{SWE-chat}~\citep{baumann2026swe}, which contains naturally occurring interactions between coding agents and developers. The resulting dataset contains 6,923 action-level instances, as summarized in Table~\ref{tab:data-source-statistics}.

\begin{table}[h]
\caption{Composition of \textsc{SWE-Intervene}.}
\label{tab:data-source-statistics}
\centering
\small
\setlength{\tabcolsep}{5pt}
\renewcommand{\arraystretch}{1.12}
\begin{tabular}{lrr}
\toprule
\textbf{Source} & \textbf{Instances} & \textbf{Proportion} \\
\midrule
\textsc{Open-SWE-Traces} & 3,451 & 49.85\% \\
\textsc{SWE-Hero}        & 1,089 & 15.73\% \\
\textsc{SWE-chat}        & 2,383 & 34.42\% \\
\midrule
\textbf{Total}           & \textbf{6,923} & \textbf{100\%} \\
\bottomrule
\end{tabular}
\end{table}

The unit of annotation is a decision point immediately before the execution of a proposed action. Each instance contains the task description \(u\), the relevant trajectory prefix \(h_t\), the complete proposed action \(a_t\), and an intervention label \(y_t\). The recorded continuation and task outcome are retained as privileged evidence for annotation and teacher training, but are excluded from the causal input available to the deployed sentinel.

Rather than labeling every step uniformly, we prioritize decision-critical nodes. In failed trajectories, these include points at which an error begins to propagate, the agent overlooks decisive evidence, or intervention could prevent an incorrect completion claim. We also retain \textsc{Allow} instances in which the agent is making reasonable progress, including imperfect actions from which it is already recovering autonomously. This distinction prevents the dataset from reducing intervention detection to local error detection.

\subsubsection{Annotation Protocol}

GPT-5.5~\citep{openai2026gpt55} assigns one of the three labels defined in Section~\ref{sec:problem-formulation}, following criteria and annotation guidelines written by human researchers. The annotator is shown the task, relevant history, complete proposed action, and recorded continuation. It is explicitly instructed to judge whether intervention at the current node would improve the probability of eventual task completion relative to allowing the agent to continue, rather than whether the action could merely be made more standardized or efficient.

For every candidate intervention, the annotation process considers three questions:

\begin{enumerate}
    \item Does the identified problem materially threaten task completion, rather than merely reflect an imperfect or nonstandard operation?
    \item Has the agent already recognized the problem and begun a reasonable recovery process?
    \item What concrete loss would an intervention prevent, or what current obstacle would it remove, compared with autonomous continuation?
\end{enumerate}

A node is labeled \textsc{Redirect} only when there is concrete evidence that the proposed action may impede task completion and corrective guidance can enable autonomous recovery. A node is labeled \textsc{Hard-Pause} when progress depends on task-specific information, authorization, or preferences unavailable to the agent. Otherwise, including cases in which the agent is already correcting its own mistake, the node is labeled \textsc{Allow}.

\subsubsection{Reconstructing Natural \textsc{Hard-Pause} Instances}

Naturally occurring \textsc{Hard-Pause} events are rare in standard coding-agent trajectories. We therefore extract additional examples from \textsc{SWE-chat} interactions containing \texttt{AskUserQuestion} (AUQ) calls. These interactions provide direct evidence that the agent encountered information, authorization, or a user preference that it could not infer autonomously.

For each retained AUQ interaction, we reconstruct the decision point immediately before the question is issued. The current AUQ call and its corresponding human response are removed from the causal input, so the sentinel must infer the need for assistance solely from information available at that decision point. Earlier AUQ exchanges may remain in the history only when they occurred strictly before the reconstructed decision point; their responses are therefore part of the legitimately observed state rather than privileged future information. Multiple valid \textsc{Hard-Pause} nodes may be extracted from one trajectory, but all nodes from the same trajectory are assigned to the same data split. We perform splitting and de-duplication at the trajectory level, yielding zero trajectory overlap between training and held-out evaluation data.

We retain only AUQ cases in which the requested information is necessary for current progress. Questions that are optional, premature, already answered by the task context, or unrelated to a present obstacle are excluded. Because the suffix following an AUQ call is conditioned on the human response, reconstructed AUQ instances provide only causal three-class supervision for the student and are never used to construct privileged-future targets for the teacher.

\subsubsection{Two-Stage Quality Audit}

We conduct a two-stage human audit focused on provisional \textsc{Redirect} and \textsc{Hard-Pause} instances, since unnecessary intervention can disrupt an otherwise successful trajectory. In the first stage, reviewers inspect the complete task context, proposed action, existing label, and recorded continuation using the three criteria above. The original annotation is retained when it is well supported; revision is required only when the trajectory provides clear evidence that the intervention is unnecessary or temporally misplaced.

The second stage independently cross-checks proposed removals and ambiguous cases. Particular attention is paid to nodes involving superficially suspicious operations, such as commands that obscure exit codes, repeated searches, or repeated builds. Such patterns are not treated as intervention-worthy by themselves: they are retained only when the surrounding task state establishes a concrete threat to completion. Semantic judgments are made from the trajectory content; automatic scripts are used only for deterministic extraction and schema validation, not for assigning or revising labels.

Among 2,252 provisional intervention candidates, the audit removes 570 instances that lack sufficient evidence for intervention, corresponding to a rejection rate of 25.3\%. The remaining 1,682 candidates exhibit a concrete completion-relevant failure risk or an obstacle that cannot be resolved without external assistance.

\begin{table}[t]
\caption{Results of the intervention-candidate audit.}
\label{tab:intervention-audit}
\centering
\small
\setlength{\tabcolsep}{7pt}
\renewcommand{\arraystretch}{1.12}
\begin{tabular}{lr}
\toprule
\textbf{Audit outcome} & \textbf{Instances} \\
\midrule
Provisional intervention candidates & 2,252 \\
Retained after audit                 & 1,682 \\
Rejected after audit                 & 570 \\
\midrule
Rejection rate                       & 25.3\% \\
\bottomrule
\end{tabular}
\end{table}

The most common rejection cases involve locally imperfect but recoverable actions, problems that the agent has already recognized and is actively addressing, and \textsc{Hard-Pause} labels placed before external information is actually required. This audit aligns the dataset with the central objective of \textsc{HiSentinel}: intervening when doing so is likely to improve final task completion, rather than correcting every detectable imperfection.


\subsection{Training details.}
We use the same training hyperparameters for the 0.6B and 1.7B
\textsc{HiSentinel} models unless otherwise specified. Training consists of
causal routing supervision with future-view distillation, feedback SFT, and
feedback preference optimization. The routing model is kept fixed during
feedback optimization. Table~\ref{tab:training-hyperparameters} summarizes
the complete configuration.

\begin{table}[t]
\centering
\caption{\textbf{Training hyperparameters of \textsc{HiSentinel}.}
The settings are shared by the Qwen3-0.6B and Qwen3-1.7B variants.
The feedback learning rate decays to its floor after the first epoch.}
\label{tab:training-hyperparameters}
\small
\setlength{\tabcolsep}{4pt}
\renewcommand{\arraystretch}{1.05}
\begin{tabularx}{\linewidth}{@{}lX@{}}
\toprule
\bfseries Hyperparameter & \bfseries Value \\
\midrule
\multicolumn{2}{@{}l}{\itshape General configuration} \\
Backbones
& Qwen3-0.6B and Qwen3-1.7B \\
Future-view teacher
& Frozen Qwen3-Coder-30B-A3B-Instruct \\
Numerical precision
& BF16 \\
Optimizer
& AdamW, $\beta_1=0.9$, $\beta_2=0.95$ \\
Weight decay
& 0.01 \\
Gradient clipping
& 1.0 \\
LoRA configuration
& Rank 32, $\alpha=64$, dropout $=0.05$ \\
LoRA target modules
& \texttt{q\_proj}, \texttt{k\_proj}, \texttt{v\_proj}, and
  \texttt{o\_proj} \\
Maximum causal input length
& 16,384 tokens \\
\addlinespace[2pt]

\multicolumn{2}{@{}l}{\itshape Causal routing and future-view distillation} \\
Training records
& 5,680 \\
Epochs / optimizer steps
& 2 / 710 \\
Effective batch size
& 16 \\
Epoch seeds
& 42 and 43 \\
Learning rate
& $5\times10^{-5}$ \\
LR schedule
& 12-step linear warmup followed by cosine decay \\
Routing objective
& Three-class route-token CE over
  \textsc{Allow}, \textsc{Redirect}, and \textsc{Hard-Pause} \\
Class-prior adjustment
& $\tau=0.85$ \\
Distillation objective
& Two-view future-to-causal binary-logit distillation \\
KD temperature / maximum weight
& $T=2.0$ / $\lambda_{\max}=0.25$ \\
KD-weight warmup
& 20 optimizer steps \\
\addlinespace[2pt]

\multicolumn{2}{@{}l}{\itshape Feedback supervised fine-tuning} \\
Training records
& 1,345 \\
Epochs / optimizer steps
& 5 / 425 \\
Batch size
& 16 \\
Epoch seeds
& 42, 43, 44, 45, and 46 \\
Learning rate
& $1\times10^{-4}$, decayed to a floor of $1\times10^{-5}$ \\
LR schedule
& 5-step linear warmup followed by cosine decay \\
Objective
& Mean next-token CE over feedback tokens and EOS \\
Route-token loss
& None; routing parameters remain fixed \\
\addlinespace[2pt]

\multicolumn{2}{@{}l}{\itshape Feedback preference optimization} \\
Training pairs
& 800 \\
Optimizer steps / maximum batch size
& 50 / 16 pairs \\
Random seed
& 42 \\
Learning rate
& $5\times10^{-6}$ \\
LR schedule
& 5-step linear warmup followed by cosine decay \\
DPO coefficient
& $\beta=0.1$ \\
Loss weights
& Preferred-response NLL $=1.0$; DPO loss $=1.0$ \\
Reference policy
& Frozen feedback-SFT checkpoint \\
Dropout
& Disabled during preference optimization \\
\bottomrule
\end{tabularx}
\end{table}

\subsection{Evaluation-Time \textsc{Hard-Pause} Protocol}
\label{app:hard-pause-protocol}

Following the evaluation design of \textsc{Ask-or-Assume} and established protocols for agents operating under missing or underspecified information, including \textsc{AskBench}, \textsc{ClarQ-LLM}, \textsc{ToolSandbox}, and \textsc{QuestBench}~\citep{edwards2026ask, zhao2026and, gan2026clarq4llm, lu2025toolsandbox, li2026questbench}, we implement \textsc{Hard-Pause} as a controlled interaction with a proxy user rather than granting the evaluated agent direct access to the complete task specification. The evaluated agent initially observes only the underspecified task input. When the sentinel emits \textsc{Hard-Pause}, execution is suspended and the agent's clarification question is forwarded to a proxy-user API.

Each proxy-user request contains only three components: a fixed role instruction defining the information boundary, the agent's exact clarification question, and the withheld complete task specification as private evidence available exclusively to the proxy user. The proxy API is not given the reference solution or patch, hidden tests, verifier outputs, evaluation labels, sentinel route or risk score, future trajectory, or any other indication of the action expected from the evaluated agent. Consequently, the complete specification cannot enter the agent context except through an admissible answer to a clarification question.

The proxy user is instructed to return only the minimum information that directly answers the submitted question and is explicitly supported by the withheld specification. It must not provide chain-of-thought reasoning, implementation advice, code, patches, test outcomes, unsolicited requirements, additional hints, or information merely inferred from the intended solution. If the requested information is not present in the withheld specification, the proxy must abstain rather than speculate. This knowledge-boundary restriction follows the general provider-agent practice used in missing-information benchmarks, in which the simulated user possesses private task facts but is prevented from revealing a complete solution.

The resulting clarification is inserted verbatim into the interaction as the user's response, after which execution resumes from the paused state. Each admitted \textsc{Hard-Pause} produces at most one proxy-user response and remains subject to the same fixed intervention budget and cooldown used throughout evaluation. We log every query and response for auditing. Proxy-user latency and token usage are reported separately and are excluded from the evaluated model's inference cost, while the wall-clock task time includes the complete interaction.

The same proxy-user API is available to every baseline in all end-to-end benchmarks, under the information boundary and response restrictions described above. Baselines may request clarification when they determine that human input is needed; access to the API is not exclusive to \textsc{HiSentinel}. In our runs, the other baselines rarely invoked it, consistent with their lack of training to recognize missing information and initiate clarification.

\end{document}